\documentstyle[11pt]{article}
\newcommand{\be}{\begin{equation}}
\newcommand{\ee}{\end{equation}}
\newcommand{\lsapprox}{\ \raise .6ex\hbox{$<$} \hskip -.15in\lower
.5ex\hbox{$\sim$}\ }
\newcommand{\grapprox}{\ \raise .6ex\hbox{$>$} \hskip -.15in\lower
.5ex\hbox{$\sim$}\ }
\begin{document}
\begin{center}
{\bf Normal-mode-based analysis of electron plasma waves \\
with second-order Hermitian formalism} \\
J.J. Ramos and R.L. White \\
Plasma Science and Fusion Center, Massachusetts Institute of Technology \\
Cambridge MA 02139 , U.S.A.  \\
\bigskip
\bigskip
\bigskip
{\bf Abstract}  \\
\end{center}
\par The classic problem of the dynamic evolution of Langmuir electron
waves in a collisionless plasma and their Landau damping is cast as a
second-order, self-adjoint problem with a continuum spectrum of real 
and positive
squared frequencies. The corresponding complete basis of singular 
normal modes is
obtained, along with their orthogonality relation. This yields easily 
the general
expression of the time-reversal-invariant solution for any 
initial-value problem.
An example is given for a specific initial condition that illustrates 
the Landau
damping of the macroscopic moments of the perturbation. \\
\
\\
\
\\
\
\\
\
\\
\
\\
\
\\
\
\\
\
\newpage

\par The linear Landau damping of collisionless plasma waves is one 
of the classic
results in plasma physics. After Landau's original formulation$^1$, 
based on the
Laplace transform of initial-value solutions for high-frequency electron
Langmuir waves, an equivalent formulation based on the normal modes 
of such a system
was developed by Van Kampen$^2$ and Case$^3$. Despite its 
attractiveness, the Van
Kampen-Case normal-mode approach is mathematically cumbersome because 
the considered
normal modes are not eigenfunctions of a Hermitian operator. A recent 
work$^4$ has
formulated the linear theory of low-frequency, collisionless sound waves in a
quasineutral plasma, in terms of the complete basis of eigenfunctions 
of a Hermitian
operator. This paper develops the analogous Hermitian (self-adjoint) 
formalism for the
normal modes of the classic, high-frequency electron plasma wave 
problem, which results
in a transparent and mathematically straightforward analysis of the 
dynamics of such
waves and their eventual Landau damping. \\
\par Consider a small-amplitude, electrostatic (${\bf B}=0 , \ \nabla 
\times {\bf E}=0$)
perturbation about a homogeneous and Maxwellian plasma equilibrium 
with immobile
ions. A linear analysis of such a perturbation can be based on the 
independent study
of uncoupled spatial-plane-wave Fourier modes characterized by their wavevector
$\bf k$. Then, for one such $\bf k$-mode, the curl-free condition on 
the electric
field implies that $\bf E$ is in the direction of $\bf k$ (${\bf E} = 
E{\bf k}/k$) and
the linearized electron Vlasov-Maxwell system yields
\be
\frac{1}{c^2} \frac{\partial E(t)}{\partial t}=-j[f_1]
\ee
\be
j[f_1]=-e\int_{-\infty}^{\infty} dv \ v f_1(v,t)
\ee
\be
\frac{\partial f_1(v,t)}{\partial t}+ikv f_1(v,t)+\frac{eE(t)v}{T_0} 
f_{M0}(v^2)=0.
\ee
Here, $j(t)$ is the magnitude of the electric current which is also parallel to
$\bf k$ (${\bf j} = j{\bf k}/k$), $v$ is the phase-space velocity component in
the direction of $\bf k$, and $f_{M0}(v^2)$ and $f_1(v,t)$ stand 
respectively for the
electron equilibrium and perturbation distribution functions, 
integrated over the
phase-space velocity components perpendicular to $\bf k$. Thus, the 
one-dimensional
Maxwellian equilibrium distribution function is
\be
f_{M0}(v^2) = n_0 \left(\frac{m}{2\pi T_0}\right)^{1/2}
\exp \left(-\frac{m v^2}{2 T_0} \right)
\ee
where $n_0$ and $T_0$ are the electron equilibrium density and temperature, and
$m$ is the electron mass. The density moment of (3) yields the 
continuity equation
\be
\frac{\partial \sigma (t)}{\partial t} + i k j(t) = 0
\ee
where
\be
\sigma (t)=-e\int_{-\infty}^{\infty} dv \ f_1(v,t)
\ee
is the charge density. Then (1) and (5) guarantee that Gauss' law,
$ikE(t)=c^2 \sigma(t)$, is satisfied at all times provided it is 
satisfied by the
initial condition at $t=0$. \\
\par Writing $f_1$ as the sum of its even and odd parts with respect to $v$
($f_1=f_1^{even}+f_1^{odd}$) and eliminating $f_1^{even}$ and $E$, 
the linearized
Vlasov-Maxwell system (1-3) reduces to the following second-order 
linear problem with
respect to time for $f_1^{odd}(v,t)$:
\be
- \ \frac{1}{k^2} \ \frac{\partial^2 f_1^{odd}}{\partial t^2}
= L[f_1^{odd}]
\ee
where the linear operator $L$ is
\be
L[f_1^{odd}] = v^2 f_1^{odd} - \frac{ec^2}{k^2T_0} j[f_1^{odd}] v f_{M0} .
\ee
The operator $L$ is self-adjoint in the Hilbert space of 
square-integrable distribution
functions with the scalar product
\be
\left\langle f | f' \right\rangle =
\int_{-\infty}^\infty dv \ \frac{T_0}{f_{M0}(v^2)} f^*(v) f'(v),
\ee
because the scalar product $\langle f | L[f'] \rangle$ can be cast in the
Hermite-symmetric form
\be
\langle f | L[f'] \rangle = \frac{c^2}{k^2} j[f^*] j[f'] +
\int_{- \infty}^\infty dv \ \frac{T_0}{f_{M0}(v^2)} v^2 f^*(v) f'(v)
= \langle L[f] | f' \rangle.
\ee
Besides,
\be
\langle f | L[f] \rangle = \frac{c^2}{k^2} |j[f]|^2 +
\int_{- \infty}^\infty dv \ \frac{T_0}{f_{M0}(v^2)} v^2 |f(v)|^2 > 0
\ee
so $L$ is a positive operator. \\
\par The normal modes of the second-order problem (7) are separable
solutions of the form
\be
f_1^{odd}(v,t) \ = v \ h^\lambda (v^2) \ \exp(-i \omega t)
\ee
where the label $\lambda$ is the squared phase velocity
($\lambda \equiv \omega^2 / k^2$) so that $vh^\lambda$ is an eigenfunction
of the operator $L$ with eigenvalue $\lambda$. Since $L$ is 
self-adjoint and positive,
the $\lambda$ spectrum is real and positive, therefore the 
normal-mode frequencies
$\omega$ are real. Then, calling $\zeta \equiv v^2$ and normalizing 
$h^\lambda$ to
\be
-\frac{1}{e} j[vh^\lambda] = \int_0^\infty d\zeta \ \zeta^{1/2} 
h^\lambda (\zeta) = 1 ,
\ee
the normal-mode eigenvalue equation can be expressed as
\be
(\zeta - \lambda) h^\lambda(\zeta) = - \frac{m \omega_p^2}{k^2 T_0 
n_0} f_{M0}(\zeta)
\ee
where $\omega_p^2 \equiv c^2 e^2 n_0 / m$ is the square of the plasma 
frequency.
For any $\lambda > 0$, this has the singular solution
\be
h^\lambda(\zeta) = - \frac{m \omega_p^2}{k^2 T_0 n_0} \
{\mathcal P} \frac{f_{M0}(\zeta)}{(\zeta - \lambda)}
+ \Lambda(\lambda) \ \lambda^{-1/2} \delta(\zeta - \lambda)
\ee
where ${\mathcal P}$ stands for the Cauchy principal value and $\delta$ is the
Dirac distribution. The coefficient $\Lambda(\lambda)$ is specified 
by the condition
that $h^\lambda$ satisfy the normalization condition (13). This yields
\be
\Lambda(\lambda) = 1 + \frac{m \omega_p^2}{k^2 T_0} \ W({\hat \lambda})
\ee
where ${\hat \lambda} \equiv m \lambda (2 T_0)^{-1}$ is the ratio of 
the squared phase
velocity to the squared electron thermal velocity, and
\be
W(\hat \lambda) \ \equiv \ \pi^{-1/2}
\int_0^\infty d\hat \zeta \ \hat \zeta^{1/2} \
{\mathcal P} \frac{\exp(-\hat \zeta)}{\hat \zeta - \hat \lambda}
\ee
which has the asymptotic behavior
$W({\hat \lambda} \rightarrow \infty) = - (2{\hat \lambda})^{-1}$. \\
\par The scalar products among these normal modes are
\be
\int_0^\infty d\zeta \ \zeta^{1/2} \frac{T_0}{f_{M0}(\zeta)}
h^\lambda(\zeta) h^{\lambda'}(\zeta)  =
\frac{T_0}{\lambda^{1/2} f_{M0}(\lambda)} D(\lambda) \delta(\lambda-\lambda')
\ee
where
\be
D(\lambda) = \left[1 + \frac{m \omega_p^2}{k^2 T_0} W({\hat \lambda}) 
\right]^2 +
\pi^2 \left(\frac{m \omega_p^2}{k^2 T_0}\right)^2
\frac{\lambda f_{M0}^2(\lambda)}{n_0^2}  ,
\ee
so normal modes with different $\lambda$ eigenvalues are orthogonal as expected
from the self-adjointness of the operator $L$. \\
\par Once the normal modes (12,15-17) of the electron plasma wave system have
been obtained, one can readily solve for any initial-value problem.
The normal modes $vh^\lambda(v^2)$ constitute a complete continuum
basis in the space of odd, square-integrable functions with the 
scalar product (9),
because they are singular eigenfunctions of a self-adjoint operator. Therefore,
any initial conditions for $f_1^{odd}$ can be expanded as
\be
f_1^{odd}(v,0) = v \int_0^\infty d \lambda \
\lambda^{-1/2} C(\lambda) \ h^\lambda(v^2)
\ee
\be
\frac{\partial f_1^{odd}(v,0)}{\partial t} = kv
\int_0^\infty d \lambda \ S(\lambda) \ h^\lambda(v^2).
\ee
Then, the solution of the corresponding initial-value problem is
\be
f_1^{odd}(v,t) = v \int_0^\infty d \lambda \
\lambda^{-1/2} h^\lambda(v^2)
\Big[C(\lambda) \cos(\lambda^{1/2} k_\parallel t) +
S(\lambda) \sin(\lambda^{1/2} k_\parallel t) \Big] .
\ee
Recalling the normalization condition (13) and changing the 
integration variable
back to $\omega$, one obtains the expression for the current perturbation
\be
j(t) = - \frac{2e}{k} \int_0^\infty d\omega \
\left[C\Big(\frac{\omega^2}{k^2}\Big) \cos \omega t +
S\Big(\frac{\omega^2}{k^2}\Big) \sin \omega t \right]
\ee
which means that, up to the multiplicative constant specified in Eq.(23),
$C(\omega^2 / k^2)$ is the cosine Fourier transform of $j(t)$ and
$S(\omega^2 / k^2)$ is its sine Fourier transform. From Maxwell's equation (1)
and Gauss's law or the continuity equation (5), the electric field 
and the electron
density perturbation are
\be
E(t) = \frac{2ec^2}{k} \int_0^\infty \frac{d\omega}{\omega} \
\left[C\Big(\frac{\omega^2}{k^2}\Big) \sin \omega t -
S\Big(\frac{\omega^2}{k^2}\Big) \cos \omega t \right]
\ee
\be
n_1(t) = - \frac{\sigma(t)}{e} = -2i \int_0^\infty \frac{d\omega}{\omega} \
\left[C\Big(\frac{\omega^2}{k^2}\Big) \sin \omega t -
S\Big(\frac{\omega^2}{k^2}\Big) \cos \omega t \right]
\ee
This solution exhibits the invariance under the time reversal,
\be
t \rightarrow - t, \quad
f_1^{odd}(v,0) \rightarrow - f_1^{odd}(v,0), \quad
\frac{\partial f_1^{odd}(v,0)}{\partial t}  \rightarrow
\frac{\partial f_1^{odd}(v,0)}{\partial t},
\ee
and the Landau damping of the macroscopic variables for $t 
\rightarrow \pm \infty$,
as the consequence of the superposition of a continuum of spectral 
components with
rapidly varying phases. This analysis provides also a simple linear 
proof that any
$j(t)$, $E(t)$ or $n_1(t)$, such that its Fourier transform exists, 
can be realized
with the appropriately chosen initial condition defined explicitly by
Eqs.(20,21,23-25). This is a result that was originally proven non-linearly by
Bernstein, Greene and Kruskal$^5$ and whose linear limit was not always well
understood. \\
\par An explicit application of the above formalism is given next, by 
considering the
specific initial condition
\be
f_1(v,0) = \frac{n_1(0)}{n_0} f_{M0}(v^2), \quad E(0)=\frac{iec^2}{k}n_1(0) .
\ee
This implies
\be
f_1^{odd}(v,0) = 0
\ee
and, from Eq.(3),
\be
\frac{\partial f_1^{odd}(v,0)}{\partial t} = -ikv
\left(1 + \frac{m \omega_p^2}{k^2 T_0}\right) \frac{n_1(0)}{n_0} f_{M0}(v^2) .
\ee
Making use of the orthogonality relation (18,19), the projection of 
this initial
condition onto the normal-mode basis yields $C(\lambda)=0$ and
\be
S(\lambda) = -i
\left(1 + \frac{m \omega_p^2}{k^2 T_0}\right)
\frac{n_1(0)\lambda^{1/2}f_{M0}(\lambda)}{n_0D(\lambda)} .
\ee
Accordingly, the time evolution of density perturbation is given by
\be
n_1(t) = \Big( \frac{2}{\pi} \Big)^{1/2}
\int_0^\infty d\omega \ {\tilde n}(\omega) \ \cos \omega t ,
\ee
where the Fourier transform of $n_1(t)$ is
\be
{\tilde n}(\omega) = (2 \pi)^{1/2}
\left(1 + \frac{m \omega_p^2}{k^2 T_0}\right)
\frac{n_1(0) f_{M0}(\omega^2/k^2)}{n_0 k D(\omega^2/k^2)}
\ee
which can have a sharp resonant peak if
$D(\omega^2/k^2)$ becomes close to zero for a narrow frequency interval. This
happens if and only if the wave phase velocity is much greater than 
the electron
thermal velocity, i.e. $\omega^2/k^2 \gg 2T_0/m = v_{th}^2$, so that
$[1 + m \omega_p^2 k^{-2} T_0^{-1} W({\hat \lambda})]^2$ can have a zero
with ${\hat \lambda} \gg 1 $, for which the
other positive term in the expression of $D(\lambda)$ (19) is small. 
Then, using the
large-argument asymptotic form of $W({\hat \lambda})$, one can approximate
\be
1 + \frac{m \omega_p^2}{k^2 T_0} W({\hat \lambda}) \simeq
1 - \frac{\omega_p^2}{\omega^2}
\ee
which has a zero at $\omega = \omega_p$. Substituting the
approximation (33) and setting $\omega = \omega_p$ in the remaining 
terms of (32),
in the limit $\omega_p \gg k v_{th}$, one obtains
\be
\frac{{\tilde n}(\omega)}{n_1(0)} \ \simeq \ \Big(\frac{8}{\pi}\Big)^{1/2}
\frac{\eta_0 \ \omega_p^3}{(\omega^2-\omega_p^2)^2 + 4 \eta_0^2 \ \omega_p^4}
\ee
where
\be
\eta_0  = \pi^{1/2}
\Big(\frac{\omega_p}{k v_{th}}\Big)^3
\exp\Big(-\frac{\omega_p^2}{k^2 v_{th}^2}\Big) \ll 1 .
\ee
Finally, after substituting (34) in (31) and carrying out the integration over
$\omega$, the corresponding approximation for $n_1(t)$ is
\be
\frac{n_1(t)}{n_1(0)} \simeq \exp(-\eta_0 \omega_p |t|)
\Big[ \cos(\omega_p t) + \eta_0 \sin(\omega_p |t|) \Big] ,
\ee
in agreement with the classic result$^{1,2,3}$ for the weakly Landau-damped
electron plasma wave. Consistent with the time-reversal invariance, 
$n_1(t)$ is an
even function of time with the same decaying behavior as $t 
\rightarrow \pm \infty$. \\
\
\\
\
\\
{\bf Acknowledgements} \\
\indent
This work was sponsored by the U.S. Department of Energy under Grant
No. DEFG02-91ER54109 at the Massachusetts Institute of Technology. 
One of the authors
(R.L.W.) was also supported by the U.S. Department of Energy Fusion 
Energy Sciences
Postdoctoral Reasearch Program administered by the Oak Ridge 
Institute for Science
and Education (ORISE) for the DOE. \\
\
\\
\
\\
{\bf References} \\
\indent
$^1$L. Landau, J. Phys. (U.S.S.R.) $\bf 10$, 25 (1946). \\
\indent
$^2$N.G. Van Kampen, Physica $\bf 21$, 949 (1955). \\
\indent
$^3$K.M. Case, Ann. Phys. $\bf 7$, 349 (1959). \\
\indent
$^4$J.J. Ramos, J. Plasma Phys. $\bf 83$, 725830601  (2017). \\
\indent
$^5$I.B. Bernstein, J.M. Greene and M.D. Kruskal, Phys. Rev. $\bf 
108$, 546 (1957).

\end{document}